\documentclass[10pt,conference,letterpaper]{IEEEtran}

\usepackage{silence}
\newif\ifproposal
    \proposalfalse

\newif\ifthesis
    \thesisfalse

\newif\ifieeejournal
    \ieeejournalfalse
\newif\ifacmjournal
    \acmjournalfalse
\newif\ifbiographies
    \biographiesfalse
\newif\ifreplytoreviewers
    \replytoreviewersfalse

\newif\ifieeeconference
    \ieeeconferencefalse

\newif\ifisf
    \isffalse

\newif\ifmicro
    \microfalse

    \ieeeconferencetrue

\usepackage{packages/basic_packages} 
\renewcommand{\eqref}[1]{(\ref{#1})}
\newcommand{\secref}[1]{\mbox{Section~\ref{#1}}}

\newcommand{\figref}[1]{\mbox{Fig.~\ref{#1}}}

\newcommand{\tblref}[1]{\mbox{Table~\ref{#1}}}

\definecolor{applegreen}{rgb}{0.55, 0.71, 0.0}
\newcommand{\black}[1]{{\color{black}#1}} % write text in blue
\usepackage{environ}
\newcounter{charcount}
\newcounter{charlim}
\NewEnviron{countenv}[2][]{%
  \setcounter{charcount}{0}%
  \expandafter\countem\BODY\relax\EOE
  \relax%
  \ifnum\value{charcount}<\value{charlim}\relax
      \begin{#2}\BODY\end{#2}
  \else
      \begin{#2}#1 \BODY  
         \color{red}\thecharcount{} letters in the #2 exceeds
         \thecharlim{} character limit. 
         Use the ``charlim'' parameter to change the character limit. \color{black}
       \end{#2} 
  \fi
}

\long\def\countem#1#2\EOE{%
  \stepcounter{charcount}%
  \ifx\relax#2
    \def\next{\relax}%
  \else
    \def\next{\countem#2\EOE}%
  \fi
  \expandafter\next%
}

\usepackage{scalerel}
\usetikzlibrary{svg.path}
\definecolor{orcidlogocol}{HTML}{A6CE39}
\tikzset{
  orcidlogo/.pic={
    \fill[orcidlogocol] svg{M256,128c0,70.7-57.3,128-128,128C57.3,256,0,198.7,0,128C0,57.3,57.3,0,128,0C198.7,0,256,57.3,256,128z};
    \fill[white] svg{M86.3,186.2H70.9V79.1h15.4v48.4V186.2z}
                 svg{M108.9,79.1h41.6c39.6,0,57,28.3,57,53.6c0,27.5-21.5,53.6-56.8,53.6h-41.8V79.1z M124.3,172.4h24.5c34.9,0,42.9-26.5,42.9-39.7c0-21.5-13.7-39.7-43.7-39.7h-23.7V172.4z}
                 svg{M88.7,56.8c0,5.5-4.5,10.1-10.1,10.1c-5.6,0-10.1-4.6-10.1-10.1c0-5.6,4.5-10.1,10.1-10.1C84.2,46.7,88.7,51.3,88.7,56.8z};
  }
}

\newcommand\orcidicon[1]{\href{https://orcid.org/#1}{\mbox{\scalerel*{
\begin{tikzpicture}[yscale=-1,transform shape]
\pic{orcidlogo};
\end{tikzpicture}
}{|}}}}
\newcommand{\V}{\,\si{\volt}\xspace}
\newcommand{\mV}{\,\si{\milli\volt}\xspace}

\newcommand{\mm}{\,\si{\mm}\xspace}
\newcommand{\mmsquared}{\,\si{\mm\squared}\xspace}
\newcommand{\m}{\,\si{\m}\xspace}

\newcommand{\nm}{\,\si{\nm}\xspace}

\newcommand{\um}{\,\si{\um}\xspace}
\newcommand{\umsquared}{\,\si{\um\squared}\xspace}

\newcommand{\s}{\,\si{\s}\xspace}
\newcommand{\ps}{\,\si{\ps}\xspace}
\newcommand{\ns}{\,\si{\ns}\xspace}
\newcommand{\us}{\,\si{\us}\xspace}
\newcommand{\ms}{\,\si{\ms}\xspace}

\newcommand{\bit}{\,\si{\bit}\xspace}

\newcommand{\Hz}{\,\si{\Hz}\xspace}
\newcommand{\kHz}{\,\si{\kHz}\xspace}
\newcommand{\MHz}{\,\si{\MHz}\xspace}
\newcommand{\GHz}{\,\si{\GHz}\xspace}

\newcommand{\mW}{\,\si{\milli\watt}\xspace}

\newcommand{\C}{\,\si{\celsius}\xspace}

\newacronym{iot}{IoT}{internet-of-things}

\newacronym{snr}{SNR}{signal-to-noise ratio}
    
\newacronym{ppa}{PPA}{performance, power, and area}

\newacronym{cdf}{CDF}{cumulative distribution function}
\newacronym{pdf}{PDF}{probabililty distribution function}
\newacronym{ip}{IP}{intellectual property}
\newacronym{fir}{FIR}{finite impulse response}
\newacronym{dsp}{DSP}{digital signal processing}
\newacronym{lut}{LUT}{look-up table}
\newacronym{mac}{MAC}{multiply-and-accumulate}

\newacronym{dl}{DL}{deep learning}
    
\newacronym{ml}{ML}{machine learning}
    
\newacronym{ai}{AI}{artificial-intelligence}

\newacronym{cnn}{CNN}{convolutional neural network}
    
\newacronym{dnn}{DNN}{deep neural network}

\newacronym[longplural={graphic processing units}]{gpu}{GPU}{graphic processing unit}

\newacronym[longplural={central processing units}]{cpu}{CPU}{central processing unit}

\newacronym{tpu}{TPU}{tensor processing unit}
    
\newacronym{relu}{ReLu}{rectified linear unit}
    
\newacronym{llm}{LLM}{Large Language Model}

\newacronym{enics}{EnICS}{Emerging NanoScaled Integrated Circuits \& Systems}
    
    \newcommand{\enicsAffiliation}{EnICS Labs, Faculty of Engineering, Bar Ilan University, Ramat Gan 5290002, Israel\xspace} 

\newacronym{BIU}{BIU}{Bar-Ilan University\xspace}
     
\newacronym{UNICAL}{UNICAL}{University of Calabria\xspace}
    
\newacronym{DIMES}{DIMES}{Department of Computer Engineering, Modeling, Electronics and Systems\xspace}
    
\newacronym{USFQ}{USFQ}{Universidad San Francisco de Quito\xspace}
         
\newacronym{EPFL}{EPFL}{\'Ecole Polytechnique F\'ed\'erale de Lausanne\xspace}
     
\newacronym{isf}{ISF}{Israel Science Fund\xspace}
\newacronym{iia}{IIA}{Israel Innovation Authority\xspace}

\newacronym{itrs}{ITRS}{International Technology Roadmap for Semiconductors}
\newacronym{vlsi}{VLSI}{very large scale integration}
\newacronym{asic}{ASIC}{application specific integrated circuit}
\newacronym{pcb}{PCB}{printed circuit board}
\newacronym{cmos}{CMOS}{complementary-metal-oxide-semiconductor}

\newacronym[longplural={systems-on-chip}]{soc}{SoC}{system-on-chip}
    \newcommand{\soc}{\gls{soc}\xspace}

\newacronym[longplural={integrated circuits}]{ic}{IC}{integrated circuit}

\newacronym{mc}{MC}{Monte Carlo}
     
\newacronym{mep}{MEP}{minimum energy point}

\newacronym[longplural={non-volatile memories}]{nvm}{NVM}{non-volatile memory}

\newacronym{vdd}{$V_{\text{DD}}$}{supply voltage}
    
\newacronym{gnd}{$GND$}{ground}
    
\newacronym{subvt}{sub-$V_{\text{T}}$}{sub-threshold}
    
\newacronym{nearvt}{near-$V_{\text{T}}$}{near threshold}
\newacronym{vt}{$V_{\text{T}}$}{threshold voltage}

\newacronym{vgs}{$V_{\text{GS}}$}{gate-to-source voltage} 
\newacronym{vds}{$V_{\text{DS}}$}{drain-to-source voltage} 
\newacronym{vbs}{$V_{\text{BS}}$}{body-to-source voltage} 
\newacronym{vgb}{$V_{\text{GB}}$}{gate-to-body voltage} 
\newacronym{dibl}{DIBL}{drain induced barrier lowering}
\newacronym{gidl}{GIDL}{gate induced drain leakage} 
\newacronym{ids}{$I_{\text{DS}}$}{drain-to-source current}
\newacronym{sce}{SCE}{short channel effect}
\newacronym{rsce}{RSCE}{reverse short channel effect}
\newacronym{tox}{$t_{\text{ox}}$}{gate oxide thickness}
\newacronym{L}{$L$}{channel length}
\newacronym{W}{$W$}{channel width}
\newacronym{rbb}{RBB}{reverse body biasing}
\newacronym{fbb}{FBB}{forward body biasing}
\newacronym{btbt}{BTBT}{band-to-band tunneling}
\newacronym{bjt}{BJT}{bipolar junction transistor}
\newacronym{hvt}{HVT}{high threshold voltage}
\newacronym{lvt}{LVT}{low threshold voltage}
\newacronym{nvt}{NVT}{nominal threshold voltage}
\newacronym{pmos}{PMOS}{p-type MOSFET}
\newacronym{nmos}{NMOS}{n-type MOSFET}
\newacronym{isub}{$I_\text{sub}$}{sub-threshold leakage}
\newacronym{igate}{$I_\text{gate}$}{gate leakage}
\newacronym{ibulk}{$I_\text{bulk}$}{bulk leakage}
\newacronym{vbb}{$V_{\text{BB}}$}{body voltage} 
    
\newacronym{ptm}{PTM}{predictive technology model}

\newacronym{pdk}{PDK}{process design kit}

\newcommand{\one}{`1'\xspace}
\newcommand{\zero}{`0'\xspace}
\newacronym{sc}{SC}{standard cell}
\newacronym{vtc}{VTC}{voltage transfer characteristic}
\newacronym{dff}{DFF}{Data Flip-Flop}
\newacronym{dcvsl}{DCVSL}{differential cascade voltage switch logic}

\newacronym{dif}{DIF}{digital implementation flow}
\newacronym{hdl}{HDL}{hardware description language}
\newacronym{rtl}{RTL}{register transfer level}
\newacronym{eda}{EDA}{electronic design automation}
\newacronym{cad}{CAD}{computer-aided design}
\newacronym{pr}{P\&R}{place and route}
\newacronym{cts}{CTS}{clock-tree synthesis}
\newacronym{sta}{STA}{static timing analysis}
\newacronym{edi}{EDI}{Cadence Encounter Design Implementation}
\newacronym{s_dc}{DC}{Synopsys Design Compiler}
\newacronym{sdc}{SDC}{Synopsys Design Constraints}
\newacronym{vcd}{VCD}{value change dump}
\newacronym{pvt}{PVT}{Process-Voltage-Temperature}
\newacronym{scm}{SCM}{standard cell memory}
    
\newacronym{ips}{IPS}{instructions per second}

\newacronym{eflash}{eFlash}{embedded Flash}

\newacronym[longplural={Storage Class Memories}]{scmems}{SCM}{Storage Class Memory}

\newacronym{ddr}{DDR}{dual-data rate}

\newacronym{sata}{SATA}{Serial Advanced Technology Attachment}
    
\newacronym{nvme}{NVMe}{Non-Volatile Memory Express}
    
\newacronym{pcie}{PCIe}{Peripheral Component Interconnect Express}
     
\newacronym[longplural={hard-Disk drives}]{hdd}{HDD}{hard-Disk drive}

\newacronym[longplural={solid-State drives}]{ssd}{SSD}{solid-State drive}

\newacronym[longplural={high-bandwidth memories}]{hbm}{HBM}{high-bandwidth memory}

\newacronym[longplural={dual-inline memory modules}]{dimm}{DIMM}{dual-inline memory module}

\newacronym[longplural={dynamic random-access memories}]{dram}{DRAM}{dynamic random-access memory}

\newacronym{fifo}{FIFO}{first-in first-out}
\newacronym{lifo}{LIFO}{last-in first-out}
\newacronym[longplural={content addressable memories}]{cam}{CAM}{content addressable memory}
    \newcommand{\cam}{\gls{cam}\xspace}
\newacronym{L1}{L1}{level-1}
\newacronym{L2}{L2}{level-2}
\newacronym{L3}{L3}{level-3}
\newacronym{L4}{L4}{level-4}
\newacronym{simd}{SIMD}{single-instruction multiple-data}

\newacronym{bc}{BC}{bitcell}

\newacronym{bl}{BL}{bitline}

\newacronym{sln}{SL}{sourceline}

\newacronym{wl}{WL}{wordline}

\newacronym[longplural={Gain-Cell embedded DRAMs}]{gcedram}{GC-eDRAM}{Gain-Cell embedded DRAM}

\newacronym{sixt}{6T}{6-transistor}
    
\newacronym[longplural={static random-access memories}]{sram}{SRAM}{static random-access memory}

\newacronym[longplural={six-transistor static random access memories}]{sixtsram}{6T-SRAM}{six-transistor static random access memory}
    \newcommand{\sixtsram}{\gls{sixtsram}\xspace}

\newacronym[longplural={embedded DRAMs}]{edram}{eDRAM}{embedded DRAM}

\newacronym[longplural={multi-level cells}]{mlc}{MLC}{multi-level cell}

\newacronym{mw}{MW}{write transistor}
\newacronym{mr}{MR}{read transistor}
\newacronym{sn}{SN}{storage node}

\newacronym{wwl}{WWL}{write word line}

\newacronym{rwl}{RWL}{read word line}

\newacronym{sa}{SA}{sense amplifier}
\newacronym{drv}{DRV}{data retention voltage}
\newacronym{nwl}{NWL}{negative word line}
\newacronym{bist}{BIST}{built-in self-test}
\newacronym{bisr}{BISR}{built-in self-repair}
\newacronym{ecc}{ECC}{error correction code}
\newacronym{snm}{SNM}{static noise margin}
\newacronym{rsnm}{RSNM}{read static noise margin}
\newacronym{wsnm}{WSNM}{write static noise margin}
\newacronym{dnm}{DNM}{dynamic noise margin}
\newacronym{drt}{DRT}{data retention time}

\newacronym{lrs}{LRS}{low resistance state}

\newacronym{hrs}{HRS}{high resistance state}

\newacronym[longplural={phase-change memories}]{pcm}{PCM}{phase-change memory}

\newacronym[longplural={resistive RAMs}]{rram}{RRAM}{resistive RAM}

\newacronym{stt}{STT}{spin-transfer torque}

\newacronym[longplural={spin-transfer torque magnetic random-access memories}]{sttmram}{STT-MRAM}{spin-transfer torque magnetic random-access memory}

\newacronym[longplural={magnetic random-access memories}]{mram}{MRAM}{magnetic random-access memory}

\newacronym{mtj}{MTJ}{magnetic tunnel junction}

\newacronym{smtj}{SMTJ}{single-barrier MTJ}

\newacronym{dmtj}{DMTJ}{double-barrier MTJ}

\newacronym{mim}{MIM}{metal-insulator-metal}

\newacronym{euv}{EUV}{extreme ultra-violet}
     
\newacronym{soi}{SOI}{silicon-on-insulator}
\newacronym{fdsoi}{FD-SOI}{fully-depleted silicon-on-insulator}
    
\newacronym{rdf}{RDF}{random dopant fluctuations}
\newacronym{ocv}{OCV}{on-chip variation}

\newacronym{lpa}{LPA}{Leakage Power Analysis}
\newacronym{dpa}{DPA}{Differential Power Analysis}
\newacronym{puf}{PUF}{Physical Unclonable Function}

\newacronym{ser}{SER}{soft errors}
     
\newacronym{seu}{SEU}{single-event upset}
     
\newacronym{qcrit}{$Q_\text{crit}$}{critical charge}
\newacronym{tmr}{TMR}{triple modular redundancy}
\newacronym{dmr}{DMR}{dual modular redundancy}
\newacronym{edac}{EDAC}{error detection and correction}
\newacronym{secded}{SECDED}{single error correction~-- double error detection}
\newacronym{dected}{DECTED}{double error correction~-- triple error detection}

\newacronym{smu}{SMU}{source/measure unit}
\newacronym{dmm}{DMM}{digital multimeter}
\newacronym{example}{EXMP}{\textit{Example}}

\usepackage{bm}
\usepackage[super]{nth}

\renewcommand{\eqref}[1]{(\ref{#1})}

\newcommand{\ones}{`1's\xspace}
\newcommand{\zeros}{`0's\xspace}

\newcommand{\slbar}{$\overline{SL}$\xspace} % search line bar
\newcommand{\PC}{$\overline{\text{PC}}$\xspace} 
\newcommand{\PCd}{$\overline{\text{PCd}}$\xspace}

\newcommand{\D}{$\text{D}$\xspace}

\IEEEoverridecommandlockouts
\usepackage{algorithm}
\usepackage{algpseudocode}

\title{PiC-BNN: A 128-kbit 65\,nm Processing-in-CAM-Based End-to-End Binary Neural Network Accelerator}

\author{
    \IEEEauthorblockN{
        Yuval~Harary\IEEEauthorrefmark{1},
        Almog~Sharoni\IEEEauthorrefmark{1},
        Esteban~Garzón\IEEEauthorrefmark{4},
        Marco~Lanuzza\IEEEauthorrefmark{4},
        Adam~Teman\IEEEauthorrefmark{1},
        and Leonid~Yavits\IEEEauthorrefmark{1}
    }
    \IEEEauthorblockA{\IEEEauthorrefmark{1}\enicsAffiliation}
    \IEEEauthorblockA{\IEEEauthorrefmark{4}\,Department of Computer Engineering, Modeling, Electronics, and Systems Engineering (DIMES), \\ University of Calabria, Rende, Italy}
    Email: esteban.garzon@unical.it; leonid.yavits@biu.ac.il
}

\begin{document}
\maketitle

%----------ABSTRACT-------------------%
%-------------------------------------%
\begin{abstract}
Binary Neural Networks (BNNs), where weights and activations are constrained to binary values ($+1,-1$), are a highly efficient alternative to traditional neural networks.
%BNNs are particularly well-suited for resource-constrained environments, such as edge devices and embedded systems, as they are supposed to eliminate the need for full precision (floating or fixed point) operations. 
Unfortunately, typical BNNs, while binarizing linear layers (matrix-vector multiplication), still implement other network layers (batch normalization, softmax, output layer, and sometimes the input layer of a convolutional neural network) in full precision. This limits the area and energy benefits and requires architectural support for full precision operations. We propose PiC-BNN, a true end-to-end binary in-approximate search (Hamming distance tolerant) Content Addressable Memory based BNN accelerator.
PiC-BNN is designed and manufactured in a commercial 65nm process. PiC-BNN uses Hamming distance tolerance to \black{apply the law of large numbers to} enable accurate classification without implementing full precision operations. 
PiC-BNN achieves baseline software accuracy (95.2\%) on the MNIST dataset and 93.5\% on the Hand Gesture (HG) dataset, a throughput of 560K inferences/s, and presents a power efficiency of 703M inferences/s/W when implementing a binary MLP model for MNIST/HG dataset classification.
\end{abstract}

\begin{IEEEkeywords}
Processing-in-memory, Content-Addressable Memory, Binary Neural Network, CAM, BNN, PiM.
\end{IEEEkeywords}

%----------INTRODUCTION---------------%
%-------------------------------------%
\section{Introduction}\label{sec:intro}
Associative, or \cam, is a special type of storage that allows data access by content rather than address location. 
The entire contents of \cam are searched for a query and tag the position(s) where the query matches the content, exactly or approximately, e.g., within a certain Hamming distance (HD)~\cite{garzon2025hdcam,pagiamtzisCAM2006}.
Due to their inherent parallel search capability, CAMs have been used in a variety of applications, including within the domain of artificial intelligence, where they facilitate the acceleration of neural network workloads~\cite{laguna2022hardware,halawani2019reram,choi2018content}.

%\black{An artificial neuron is typically modeled as a nonlinear function of the dot product of the neuron's weighted inputs. 
%A basic neural network comprises a series of such neurons, where weights are the trainable parameters that, in the case of a classification network, are supposed to lead to the selection of the correct (target) class. 
%Specifically, the neuron with the highest output signal level is the one pointing to the target class.}

There is a deep conceptual connection between the behavior of artificial neural networks and associative memory. 
In the former, the neuron with the highest correlation between the input (activation or feature) and the weight vectors points at the correct (target) class, because high correlation normally yields the highest output. 
In the latter, the memory row storing weights with the best (exact or most similar) match to the query (i.e., input vector) marks the target class. 
This effect is apparent in the case of binary neural networks (BNNs), where both the input activations and the weights are $+1$ or $-1$, typically coded as logic `1' and logic `0',  respectively.
Therefore, the multiplication of the weights and the input feature becomes a simple \texttt{XNOR} operation, which is also the basic computation of associative memory.
% and therefore the multiplication W$\times$X becomes XNOR

BNNs are lightweight networks that enable very high energy efficiency by replacing full-precision multiplications by one-bit \texttt{XNOR} and accumulation by \texttt{POPCOUNT}~\cite{zahedi2024bcim}. 
However, in practice, multiple network components must be implemented in full precision to avoid a significant reduction in accuracy.
%~\cite{yang2024hybnn,zhang2025fpga,zhang2021fracbnn,liu2018bi,liu2020reactnet}. 
%A series of layers are implemented using full precision floating or fixed point operations to maintain the feature representation, which otherwise degrades rather significantly. 
Such components include, among others, batch normalization and biased parametric ReLU.
%~\cite{yang2024hybnn,zhang2025fpga}. 
In a convolutional BNN, the first layer is typically implemented with full precision. %~\cite{liu2018bi,liu2020reactnet}.
Furthermore, the output (fully connected) BNN layer must be implemented with high precision
%~\cite{zhang2025fpga,zhang2021fracbnn}
to be able to confidently differentiate among classes. 
%This is common also in BNN designs that target multilayer perceptron (MLP).
%For example, for MNIST classification, Jung, \etal~\cite{jung2022crossbar} digitize the output of analog popcount to 4 through 8 bits for subsequent batch normalization and output\footnote{We use the \textit{last} and \textit{output} layer interchangeably across the paper.} layer implementation. 
%Supporting datasets with large number of classes, such as ImageNet, requires the precision of 16 bits~\cite{conti2018xnor}. 

In summary, the requirement to implement several layers of a BNN with full precision is ubiquitous and inevitable in all types of BNNs, which makes the very definition of a BNN somewhat ambiguous.  
Consequently, the energy efficiency enabled by the elimination of multiplication-accumulation is significantly degraded by the need to employ auxiliary digital units or outsource full precision layers to software execution.   

The purpose of this work is to implement a true \textit{end-to-end-binary} BNN in- and using- \cam, where all network layers are binary and implemented in-memory, with no auxiliary digital processing units and without relying on software to implement full-precision calculations. 
We present PiC-BNN, a {\cam}-based classification BNN accelerator designed and fabricated in silicon using a commercial 65\nm process. 
We evaluate our design by implementing a binary multilayer perceptron model and measuring its classification accuracy on MNIST and Hand Gesture data sets. 
The PiC-BNN design achieves (close to) full baseline software accuracy on MNIST and Hand Gesture datasets and provides very high inference throughput and energy efficiency.  
%by performing the fully connected (output) BNN layer multiple times while slightly modifying the electrical properties of the PiC-BNN \cam array. 

This work makes the following contributions:
\begin{itemize}
    \item PiC-BNN eliminates the need for full precision operations by implementing multiple executions of the output (fully connected) layer with modified electrical parameters of CAM circuitry;
    \item PiC-BNN implements said modification by adjusting the approximate search-capable CAM's Hamming distance tolerance threshold rather than changing input activations or weights;
    \item PiC-BNN was designed and manufactured in a 65\nm commercial process and evaluated using silicon measurements.
\end{itemize}

The rest of the paper is organized as follows:
\secref{sec:background} briefly presents the background on CAMs and BNNs.
\secref{sec:design} presents PiC-BNN design.
\secref{sec:Operating principle} details the operating principle of PiC-BNN.
\secref{sec:eval} presents PiC-BNN evaluation results in terms of accuracy, performance, and power.
\secref{sec:conclusion} presents the main conclusions of this work.

    \begin{figure}[t!]
     \centering
     \includegraphics[width=0.42\textwidth]{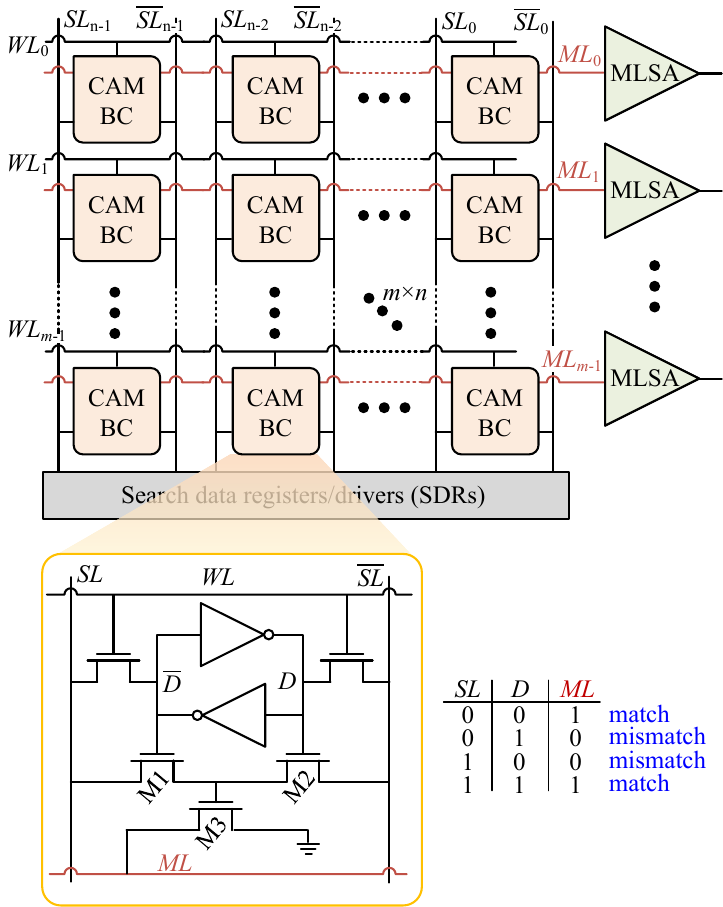}% \vspace{-3mm}
     \caption{Content-addressable memory (CAM). In the inset: SRAM-based NOR-type CAM cell and match and mismatch cases between searchline data $SL$ and stored data $D$.}
     \label{fig:cam overall architecture}
    %\vspace{-4mm}
   \end{figure}

\section{Background}\label{sec:background}
\subsection{Content Addressable Memory}\label{CAM_intro}
\figref{fig:cam overall architecture} shows the conventional CMOS NOR-type \cam array.
%with an array of $n$ columns by $m$ rows of NOR-type cells~\cite{pagiamtzisCAM2006}.
%A \cam compares between the query data held in the search data registers, and the information typically stored in \sixtsram bitcells.
The matchline ($ML$) is shared between bitcells of an $n$-bit word and also fed into a $ML$ sense amplifier (MLSA).
The searchlines ($SL$ and \slbar) are shared across all rows of the \cam array.
Read and write operations within the \cam array are executed similarly to conventional \sixtsram.
A \cam performs a comparison between the query data pattern that is driven onto the $SL$s from the search data registers (SDRs) into the memory array and the information contained within the \sixtsram bitcells ($D$ and $\overline{D}$).
%\cam read and write operations are accomplished in the same way as in the conventional \sixtsram.
%Access to a bitcell is facilitated by activating the corresponding row through the assertion of the \wl, followed by precharging to $V_{DD}$ or asserting complementary values on the SLs for read or write operations, respectively.
This search (compare) operation is done simultaneously across the entire array during a single clock cycle in two phases: (1) $ML$ precharge to $V_\text{DD}$; and (2) assertion of the query data on the $SL$s.
The MLSA evaluates the state of the $ML$ at the end of the comparison cycle and signals a match or mismatch.

In addition to exact match CMOS-based~\cite{pagiamtzisCAM2006} and emerging-memory-based~\cite{shaban2024sot,garzon2024monolithic,pan2024energy,xu2024novel} solutions, % Yavits2015Resistive,Yavits2021GIRAF,Kaplan2018PRINS,Ramanathan2020Monolithic,yin2020fecam
several CAM designs support approximate (including HD approximation), or similarity, search~\cite{ni2024tap,shaban2024sot,taha2020approximate,imani2017exploring}.
%Among these designs are soft-error tolerant CAMs, which apply error correction coding (which requires memory redundancy) and replace $ML$ sense amplifier (MLSA) with an analog comparator~\cite{Pagiamtzis2006Soft,Krishnan2009ecc}.
%Since tolerating soft errors is analogous to tolerating differences between query and stored patterns, such soft-error tolerant designs can be considered approximate search capable CAMs.
%Soft-error tolerant CAMs typically only tolerate a limited Hamming distance (HD) (up to 4 bits). 
%Another class of similarity search CAMs uses locality-sensitive hashing of stored data and query patterns~\cite{ni2019ferroelectric,riazi2017camsure}. 
%While such schemes potentially tolerate large HDs, they require the hashing of data before storage and search.
%Several emerging memory (memristor crossbar) based designs for HD approximation have also been proposed~\cite{zhu2013Hamming,taha2020approximate}. 

\begin{figure}
    \centering
\includegraphics[width=0.35\textwidth]{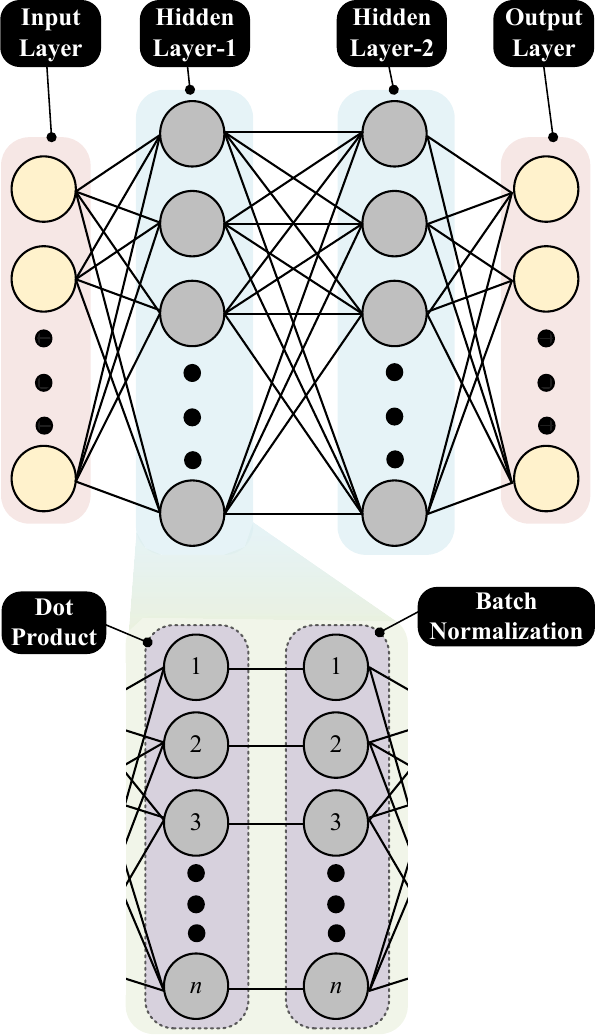}
    \caption{Binary Neural Network: multilayer perceptron (MLP) network with two hidden layers.
    Each hidden layer comprises dot product (\texttt{XNOR} and \texttt{POPCOUNT}) and batch normalization layers.}
    \label{fig:bnn_view}
\end{figure}

\subsection{Binary Neural Networks (BNNs)}
\figref{fig:bnn_view} represents a class of neural networks where weights and/or activations are constrained to binary values, typically $\{-1, +1\}$. 
This quantization significantly reduces memory consumption and computational cost, making BNNs attractive for resource-constrained environments such as embedded systems and edge devices. 
By replacing full-precision arithmetic with bitwise operations, BNNs achieve efficiency while retaining an acceptable level of predictive performance for many applications~\cite{yuan2023comprehensive}.

\begin{figure*}[t!]
    \centering
    \includegraphics[width=2.1\columnwidth]{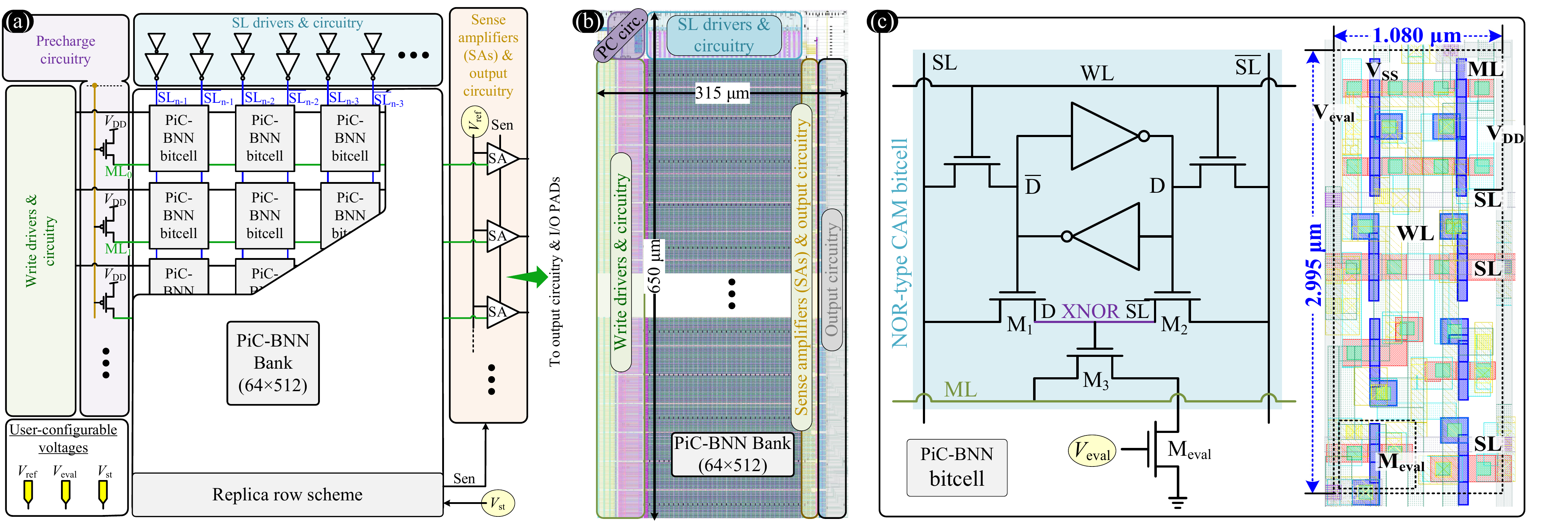}\vspace{-7mm}
    \caption{
    (a) PiC-BNN high-level architecture of a single memory bank. 
    (b) Layout of a 64$\times$512 PiC-BNN bank.
    (c) Ten-transistor PiC-BNN bitcell and bitcell layout.
    Note: User-configurable voltages are highlighted in yellow in (a) and (c).
    %The 32-kbit PiC-BNN bank is organized as two blocks along with peripherals. In the inset of (a) and (b) the ten-transistor PiC-BNN bitcell and bitcell layout are shown, respectively.
    }
    \label{fig:TOP}%\vspace{-5mm}
\end{figure*}

In the context of multilayer perceptron (MLP) models, binarization is applied to both the weight matrices and the activation functions within each layer. 
Specifically, the forward pass in an MLP with binary weights ($W$) and binary activations ($X$) is expressed as 
\begin{equation}
X_j^{l+1} = \phi\big(\texttt{sign}(\sum_iW_{ji} X_i^l + b^l))
\label{eq:bnn-act}
\end{equation}
where \( \text{sign}(\cdot) \) denotes the binarization function, \( \phi(\cdot) \) is a non-linear activation function, \black{$X_{i}^l$ is the activation of the $i$-th neuron of the $l$-th layer, $W_{ji}$ is the weight connecting the $i$-th neuron in layer $l$ to the $j$-th neuron in layer $l+1$,} and \( b^l \) represents the bias term in the \( l \)-th layer. %While this formulation simplifies the computation, it introduces non-differentiability in the \ita{backward pass}, necessitating techniques such as the straight-through estimator to approximate gradients and enable effective training.
Batch normalization is employed at each layer of the neural network~\cite{ioffe2015batch}: 

\begin{equation}
\begin{aligned}
y &= \sum_i \big( W_{ji} \cdot X_i^l \big), \\
X_j^{l+1} &= \phi \bigg(\frac{y - \mu}{\sqrt{\sigma^2 + \epsilon}} \cdot \gamma + \beta \bigg),
\end{aligned}
\label{eq:batchn}
\end{equation}
where $\mu$ and $\sigma$ are the mean and standard deviation values, and $\gamma$ and $\beta$ are trainable parameters. 
Batch normalization %shifts and scales the neuronal activations over a batch during the training process. It 
is essential in BNNs to achieve high accuracies, as it ensures that neuronal activations utilize both $+1$ and $-1$ values. 
At inference time, batch normalization manifests as a constant $C_j$ in Equation~\eqref{eq:popcount_equation}~\cite{hirtzlin2020digital}, leading to the following BNN implementation:
\begin{equation}
\text{X}_j^{l+1} = \text{sign}\left( \texttt{POPCOUNT}\left(\texttt{XNOR}(W_{ji}, X_i^l) \right) + C_j \right)
\label{eq:popcount_equation}
\end{equation}

%\black{Note that $C_j=\frac{\beta*\sqrt{\sigma^2+\epsilon}}{\gamma}-\mu$. and is derived since: 
%$$
%\frac{x-\mu}{\sqrt{\sigma^2+\epsilon}}\cdot \gamma + \beta > 0 \iff x + C_j > 0
%%$$
%}

%Despite their advantages, BNNs face challenges in representation capacity and optimization due to the extreme quantization of parameters. For MLPs, this often results in reduced expressiveness, particularly for tasks requiring fine-grained feature representations. Consequently, recent research has focused on devising training algorithms, regularization strategies, and architecture modifications to mitigate these limitations. By balancing efficiency and accuracy, BNNs offer a promising pathway for deploying neural networks in scenarios with stringent computational and energy constraints.

%More information on BNN can be found in~\cite{qin2020binary,yuan2023comprehensive}.

\subsection{Related work: hardware BNN accelerators}
%Accelerators of the first category include the following solutions.
%XOR neural engine~\cite{conti2018xnor}, XNORBIN~\cite{al2018xnorbin}, as well as a variety of later designs~\cite{nurvitadhi2016accelerating,li2022efficient,vo2022deep,ryu2023binaryware,hosseini2021binary} are digital accelerators that apply \texttt{XNOR} gates, a \texttt{POPCOUNT} network and digital accumulators to implement convolutional and fully connected binary layers. 
%FINN~\cite{umuroglu2017finn} and~\cite{zhao2017accelerating} are FPGA-based BNN accelerators supporting, among others, the CIFAR-10 dataset. 
%ReBNet~\cite{ghasemzadeh2018rebnet} represents feature maps with multiple levels of residual binarization. 
%Bi-RealNet~\cite{liu2018bi} and ReActNet~\cite{liu2020reactnet} apply full precision calculations in the first convolution layer. 
%An all-digital BNN accelerator was introduced and manufactured in~\cite{knag2020617}.
%STBNN~\cite{qiao2020stbnn} is a spatio-temporal BNN with binarized weights and activations. 
%An FPGA implementation of BNN as well as a ternary neural network is proposed in~\cite{ngadiuba2020compressing}. A structural pruning FPGA-based combined CNN and BNN accelerator was introduced in~\cite{peng2021binary}.
Since the activations and weights of BNNs are represented by 1-bit data, neural network computations can be transformed into relatively simple \texttt{XNOR} and \texttt{POPCOUNT} operations~\cite{rastegari2016xnor}.
Hardware BNN accelerators can be categorized into two main types: 
\begin{enumerate}
    \item Conventional digital accelerators that typically comprise \texttt{XNOR} gates, a \texttt{POPCOUNT} network, and digital accumulators to implement convolutional and fully connected binary layers~\cite{conti2018xnor,al2018xnorbin,nurvitadhi2016accelerating,li2022efficient,vo2022deep,ryu2023binaryware,hosseini2021binary,umuroglu2017finn,zhao2017accelerating,ghasemzadeh2018rebnet,liu2018bi,liu2020reactnet,knag2020617,qiao2020stbnn,ngadiuba2020compressing,peng2021binary}.
    \item Processing-in- or using-memory, often designed around resistive (memristive) memory~\cite{hirtzlin2020digital,jung2022crossbar,azamat2023partial,el2024afsram}. This category usually implements \texttt{POPCOUNT} in the analog domain and applies either Analog-to-Digital conversion (ADC) or Time-to-Digital conversion (TDC)~\cite{choi2018content,jung2022crossbar} in the output (fully connected) layer to preserve precision.
\end{enumerate}

%PuM BNN accelerators are typically based on emerging resistive memories and employ two of their common analog properties: 
%\begin{enumerate}
%    \item Two resistive elements (memristors) comprise \texttt{XNOR} functionality~\cite{yavits2014resistive}, which replaces a multiplier in BNN.
%    \item The crossbar column bitline accumulates the outputs of \texttt{XNOR} elements, emulating \texttt{POPCOUNT} circuit in analog domain.
%\end{enumerate}

%A ReRAM-based BNN accelerator was developed and manufactured in~\cite{hirtzlin2020digital}. The latter uses analog PuM domain to perform \texttt{XNOR} while the \texttt{POPCOUNT} is implemented using a separate digital logic block.
%An MRAM-based BNN was introduced and manufactured by~\cite{jung2022crossbar}. It uses pairs of MRAM cells to implement \texttt{XNOR}, and time-to-digital conversion to digitize the output activation. 
%Additional PuM-based BNN implementations use analog-to-digital converters (ADC) to emulate the \texttt{POPCOUNT} operation and output a full-precision output activation~\cite{azamat2023partial,el2024afsram}. 

A separate category of BNN accelerators target CAM, employing the observation that a compare (matching) operation in a CAM row is akin to a dot-product, where per-bit matching (\texttt{XNOR}) emulates single-bit multiplication and the $ML$ voltage level emulates \texttt{POPCOUNT}~\cite{choi2018content,choi2022high}.

%Such voltage level is typically digitized by applying specialized time-to-digital sensing schemes~\cite{choi2018content}.

A common disadvantage of TDC schemes is their high susceptibility to process, voltage, and temperature (PVT)  variation~\cite{leonid2022_HDCAM_Access,garzon2023low}. 
If a certain sampling time point is associated with a certain class, as in~\cite{jung2022crossbar}, a slight deviation from that point due to temperature or voltage drift may create a systematic error.
This could result in the consistent selection of an incorrect class,
%(a wrong class would be consistently selected) 
%which would be very 
making it particularly challenging to mitigate through calibration or otherwise. 

The main disadvantage of ADC-based solutions is their high area and energy overhead~\cite{shafiee2016isaac} that may render the analog processing-in or using-memory implementation inefficient compared to digital or combined implementations.

\section{PiC-BNN Design}\label{sec:design}
%PiC-BNN is an approximate search capable (HD tolerant) CAM with processing in memory capabilities. 
The PiC-BNN memory system was custom-designed in a 65\nm CMOS process, integrated within a System-on-Chip (SoC) platform, and fabricated as part of a research test chip~\cite{garzon2022risc}.
PiC-BNN comprises four 32-kbit banks, which can be logically configured as: 
\begin{itemize}
    \item 512$\times$256 array,
    \item 1024$\times$128 array, and
    \item 2048$\times$64 array.
\end{itemize}
\figref{fig:TOP}(a) shows the high-level architecture of a single PiC-BNN bank, including its peripherals: SL drivers \& circuitry, sensing circuitry, write circuitry, and precharge circuitry.
\figref{fig:TOP}(b) shows the layout of a single PiC-BNN bank, featuring an area footprint of about 0.21\mmsquared.

\figref{fig:TOP}(c) shows the PiC-BNN bitcell, made up of a conventional 9T NOR CAM cell with an extra transistor $\text{M}_\text{eval}$ in the $ML$ discharge path to ground~\cite{garzon2025hdcam,pagiamtzisCAM2006}.
From \figref{fig:TOP}(c), the PiC-BNN bitcell layout (shown at the right) presents an area footprint of about 3.24\umsquared.
%The data storage nodes of the SRAM core, \D and \Dbar,} drive the gates of a pair of nMOS transistors, \Mone and \Mtwo, which are connected to the SLs on one side, and share a common node on the other side.
%This shared node is high (\one) if either \D and \slbar are \one, or \Dbar and SL are \one. 
%In other words, it implements XNOR logic between the stored bit and the inverted query bit at the gate of \Mthree, which is connected between $ML$ and \Meval. 
%Therefore, \Mthree opens an $ML$ discharge path if a \textbf{mismatch} occurs between the query bit (SL) and the stored bit (\D).
%By controlling voltage ({$V_\text{eval}$}), $ML$ discharge pace can be adjusted, and so PiC-BNN can be configured to perform an approximate search with different (programmable) HD tolerance thresholds. 
%Lowering $V_\text{eval}$ slows down the $ML$ discharge, enabling matching input vectors whose HD from the weights stored in said row is within such HD tolerance threshold.
%The proposed PiC-BNN bitcell exhibits a compact footprint of just 3.24\umsquared, as highlighted in the inset of \figref{fig:TOP}(b).

To enable a wide range of HD tolerance in PiC-BNN search, we introduce three user-configurable voltage sources (highlighted in yellow in \figref{fig:TOP}), as follows:
\begin{itemize}
    \item MLSA reference voltage ($V_\text{ref}$): 
    By lowering (raising) the $V_\text{ref}$, we increase (reduce) the HD tolerance threshold enabled by PiC-BNN.
    \item $V_\text{eval}$: Controls the rate of PiC-BNN $ML$ discharge by adjusting the conductance of the $\text{M}_\text{eval}$ transistor. Lowering (raising) $V_\text{eval}$ slows down (speeds up) the $ML$ discharge and thus changes the HD tolerance threshold of PiC-BNN. 
    \item  $V_\text{st}$: Adjusts the sampling time of the MLSA output. By advancing (delaying) the MLSA sampling, we increase (reduce) the HD tolerance threshold enabled by PiC-BNN.
\end{itemize}

%By applying a combination of ($V_\text{ref}, V_\text{eval}, V_\text{st}$), we can tune the HD tolerance of CAM in wide range, which is required to differentiate among multiple classes with high precision. This is also the reason one 
All three user-configurable voltages are required to allow the HD tolerance to be large enough to support the required distance between the input activation and the corresponding weight vectors, as further explained in~\secref{sec:Operating principle}. 
\tblref{tab:volt} details some ($V_\text{ref}, V_\text{eval}, V_\text{st}$) combinations and the HD threshold levels enabled by them. 

\begin{table}[t!]
    \centering
    \caption{$V_\text{ref}, V_\text{eval}, V_\text{st}$ combinations and the HD tolerance threshold levels enabled by them.}
    \begin{tabular}{c|c|c|c}
        \textbf{$V$\textsubscript{ref} (\mV)} & \textbf{$V$\textsubscript{eval} (\mV)} & \textbf{$V$\textsubscript{st} (\mV)} & \textbf{HD Tolerance} \\\hline
        1200 & 1200 & 1200 & 0 \\
        750  & 950  & 1200 & 4 \\
        775  & 600  & 1200 & 8 \\
        1175 & 350  & 1150 & 12 \\
        950  & 525  & 1100 & 16 \\
        1025 & 475  & 1000 & 20 \\
        950  & 500  & 1025 & 24 \\
        775  & 600  & 1100 & 28 \\
        1175 & 400  & 1150 & 32 \\
        1000 & 475  & 725  & 36 \\
    \end{tabular}
    \label{tab:volt}
\end{table}

\section{Operating principle}\label{sec:Operating principle}
{\color{black}
PiC-BNN operating principle is based on two ideas presented below. 
The fully connected (output) layer in classification neural networks ``matches'' the flattened feature vector to a target class.
The class that yields the highest correlation with the input feature vector is supposedly the target one.  
In a CAM-based fully connected layer, such correlation manifests in a ``match degree'', as the closer the input vector to a class, i.e., the lower the HD between such vector and a weight vector representing the class, the higher the chance of such class to be the correct (target) one. 
In a BNN, the logit at the output associated with the target class should ideally be \one, while all other classes should ideally output \zeros. 
However, in real life, the target class returns \zero with some probability while some of the ``wrong'' classes accidentally return \ones. 
To mitigate this, prior BNN solutions implement \texttt{POPCOUNT} at the output of the fully connected layer at high precision by applying ADC or TDC in processing-in-memory or processing-using-memory solutions.

The law of large numbers inspires the first PiC-BNN idea: we hypothesize that if the fully connected layer is executed on the same input multiple times under different (albeit slightly) conditions, the average of the target class output will converge at \one, while the averages of unrelated classes will converge at \zero. 
In other words, the majority\footnote{It could be a simple or special majority, with a threshold higher than half.} of the target class outputs across multiple passes is very likely to be \one, while the majority of the outputs of any other class is very likely to be \zero.
}
%\black{Two distinct methods can be employed to induce such variation across passes:
%\begin{enumerate}
%    \item modifying either the input vector or the weights of the fully connected layer in each pass.
%    Therefore, creating a variable HD between the input activation vector and the weights stored in the CAM rows;
%    \item replacing the input or weights modification by applying approximate matching with \textbf{\textit{varying HD tolerance}} threshold as a way of allowing variable HD between the input vector and the weights.
%\end{enumerate}
%In this work, we adopt the second approach, i.e., varying the HD tolerance threshold of PiC-BNN.
%}
\black{Such a difference in execution conditions can be achieved by modification of either the input vector or the weights of the fully connected layer in each pass. 
The purpose of such a modification is to create a variable HD between the input activation vector and the weights stored in the CAM rows. %Such HD must be sufficient to create a different mix of matches in each pass, but not as large as to cause a consistent miss of the target class. 

Our second core idea is replacing the input or weights modification by applying approximate matching with \textbf{\textit{varying HD tolerance}} threshold as a way of allowing variable HD between the input vector and the weights.}

\begin{figure}
    \centering
\includegraphics[width=0.44\textwidth]{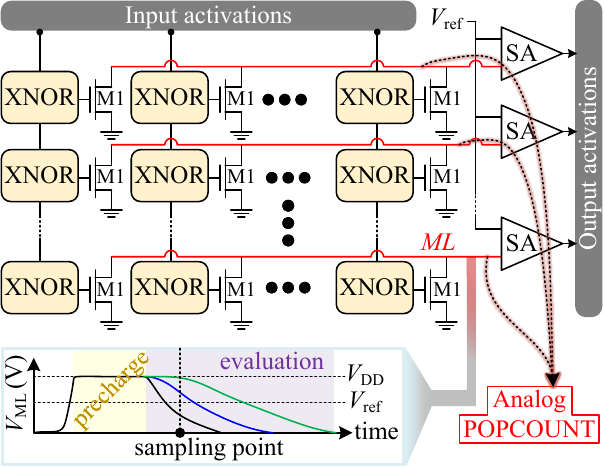}\vspace{-2mm}
    \caption{PiC-BNN concept.}
    \label{fig:cap_concept}
\end{figure}

The concept of PiC-BNN is presented in~\figref{fig:cap_concept}. 
Logically, each PiC-BNN row performs the following operation, whose components are detailed below:

\begin{equation}
\mathbf{X}_j^{l+1} = \texttt{MAJ} \left( \sum_i \texttt{XNOR}(W_{ji}, \mathbf{X}_i^l)  + \sum_i c_{ji} \right),
\label{eq:column}
\end{equation}
where \texttt{MAJ} is a logic majority operation, and $c_{ji}$ 
%$ ~(\forall i \in [0, 31], j \in [0, 63]) $ 
are the bits of the batch normalization constant of~\eqref{eq:popcount_equation} $\left(C_j=\sum_ic_{ji}\right)$.

The following details the components of \eqref{eq:column} and their implementation:
\begin{itemize}
    \item \texttt{XNOR} is implemented by a per-bit matching in a PiC-BNN cell that stores a binary weight $W_{ji}$. 
    The binary input activation $X_i$ bit is asserted on $SL$ (\slbar). 
    If $X_i \neq W_{ji}$ (representing $-1$), the $ML$ discharge path through the cell opens. 
    Otherwise, if $X_i=W_{ji}$ (representing $+1$), the $ML$ does not discharge.
    
    \item Batch normalization parameters ($\beta, \gamma, \mu, \sigma$ of \eqref{eq:batchn}) are either trainable or calculated during inference, and thereby are known in advance.
    Therefore, it is possible to replace batch normalization by adding to the output of a network layer (a dot product) a batch normalization constant $C_j$ as presented in \eqref{eq:popcount_equation}.
    %(refer to~\figref{fig:tall_illustration}\black{(a)}). 
    %\vspace{2mm}
    
    In PiC-BNN, we use the same memory to calculate the binary dot product and to add to it the batch normalization constant $C_j$ represented as a series of $+1$s and $-1$s whose sum equals $C_j$. 
    For example, $C_j=+12$ is represented by 12 matching CAM cells.
    Batch normalization is therefore implemented by adding a number of $+1$s or $-1$s %($c_{ji}$) which amounts to the batch normalization constant $C_j=\sum_ic_{ji}$, 
    to the $\sum_i \texttt{XNOR}(W_{ji}, X_i^l)$. 
    
    \item Majority is implemented as follows: $ML$ is precharged (refer to $ML$ inset in \figref{fig:cap_concept}) and the input activations $X^l$
    %$X^l_i$($i \in [0, 63]$) 
    are asserted. 
    The MLSA reference voltage $V_\text{ref}$ is calibrated such that the $ML$ voltage ($V_{ML}$) crosses it at the sampling time when the number of matching and mismatching bitcells in a row is the same.
    This way, if the number of matches exceeds the number of mismatches (refer to green line at the sampling point in~\figref{fig:cap_concept}), the $ML$ discharges slowly and so $V_{ML}$ crosses $V_\text{ref}$ after the MLSA sampling, generating $X^{l+1}_i=1$ (representing $+1$). 
    Otherwise, if the number of matching cells is below the number of mismatching ones (refer to black line at the sampling point in~\figref{fig:cap_concept}), the $ML$ discharges faster, and $V_{ML}$ crosses the $V_\text{ref}$ before the MLSA sampling time, signaling $X^{l+1}_i=0$ (representing $-1$).
    
    \item To change the execution conditions of the fully connected (output) layer, we regulate the $V_\text{ref}$, adjust the pace of $ML$ discharge by controlling $V_\text{eval}$, and set the MLSA sampling time by tuning $V_\text{st}$. 
    Adjusting these three voltages, we induce changes in the HD between the input vector and the weights associated with a certain class, which creates a sufficient difference in each fully connected layer execution as to obtain slightly different outputs in each such execution.   
\end{itemize}

\section{Evaluation}\label{sec:eval}
\subsection{Accuracy evaluation}\label{sec:eval_setup}
We evaluate the accuracy of the BNN MLP implemented in PiC-BNN using: 
\begin{itemize}
    \item The MNIST dataset, with 10 image classes and an image size of $28\times28=784$.
    \item The Hand Gesture recognition dataset~\cite{rishabh2021handgesture}, with 20 classes and an image size of $64\times64=4096$.
\end{itemize}

%We develop and train the following models:
%\begin{itemize}
 %   \item Deep model (three hidden layers): $784 \rightarrow 4096 \rightarrow 4096 \rightarrow 128 \rightarrow 10$. This model was developed in~\cite{hubara2016binarized} to provide the highest BNN MLP accuracy achieved for the MNIST dataset.
  %  \item Shallow model (no hidden layers): $784 \rightarrow 128 \rightarrow 10$.
%\end{itemize}

We develop and train the following MLP models
\begin{itemize}
    \item $784 \rightarrow 128 \rightarrow 10$
    \item $4096 \rightarrow 128 \rightarrow 20$
\end{itemize}
To evaluate PiC-BNN classification accuracy, we conduct the following experiment (Algorithm~\ref{alg:hd_inference}): 
%For each image in each test data set, the input layer $784 \rightarrow 128$ is executed once. Afterwards, the output layer $128 \rightarrow 10$ is executed several times as follows: 

\begin{algorithm}
\caption{Inference procedure with variable HD tolerance}
\label{alg:hd_inference}
\begin{algorithmic}[1]
\For{each image in each test dataset}
    \State Execute input layer: $784\,\,(4096) \rightarrow 128$
    \For{$\text{HD threshold} \in \{0, 2, 4, \dots, 64\}$}
        \State Execute output layer: $128 \rightarrow 10\,\,(20)$
        \State Log the output
    \EndFor
    \State Determine final prediction by majority vote in each class over the 33 individual outputs
\EndFor
\end{algorithmic}
\end{algorithm}

\begin{figure}[!t] % [b]-> bottom, [t]->top, [H]->Here! ([h!] should do a better job), {figure*}->float
     \centering
     \includegraphics[width=0.99\columnwidth]{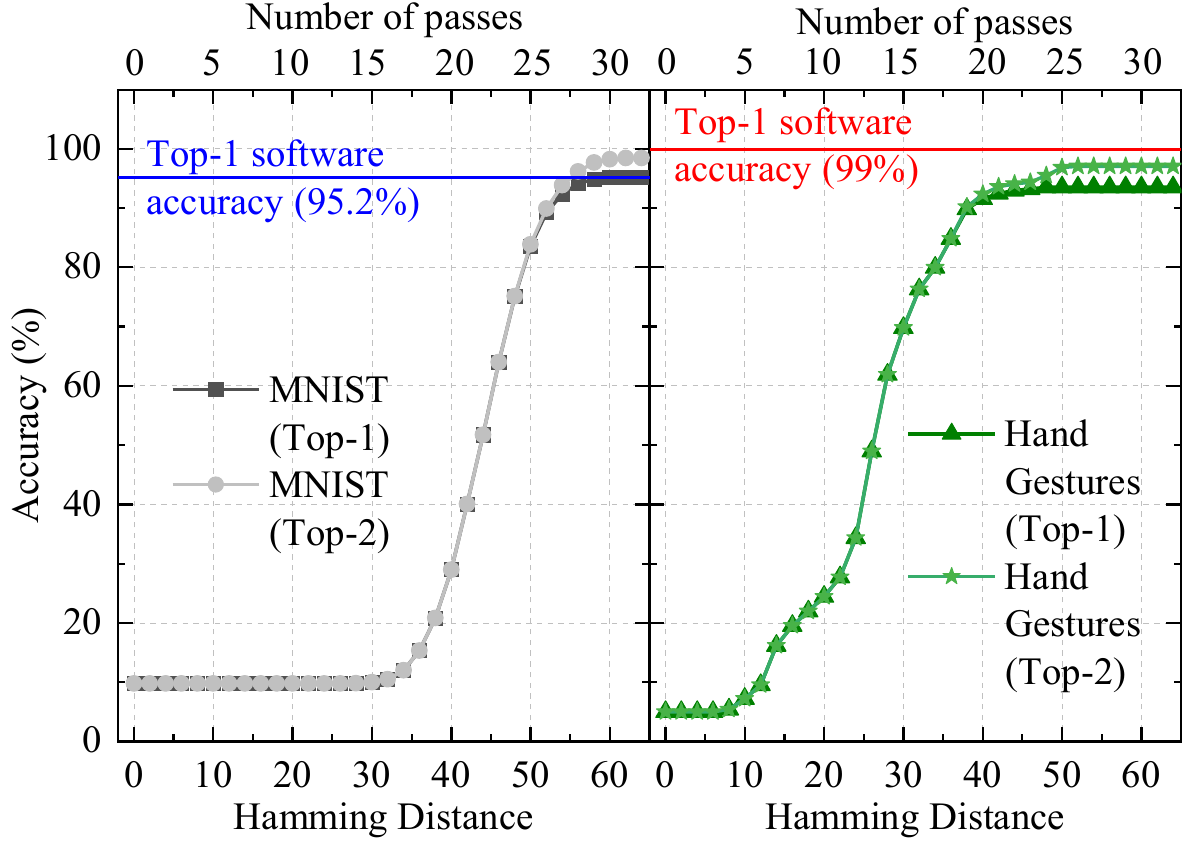} \vspace{-5mm}
     \caption{TOP-1 and TOP-2 accuracy results for MNIST and Hand Gesture datasets. 
     Note: X-axes are the number of fully connected (output) layer executions and corresponding HD tolerance threshold levels, located at the top and bottom axes, respectively. 
     }%\vspace{-5mm}
     \label{fig:accuracy_results}
\end{figure}

Accuracy evaluation results are presented in~\figref{fig:accuracy_results}. 
The accuracy grows with the number of output $128 \rightarrow 10\,\,(20)$ layer executions and correspondingly growing HD tolerance threshold.
Therefore, the PiC-BNN accuracy reaches the baseline software accuracy for MNIST (Top-1 95.2\%) and Top-1 accuracy of 93.5\% for Hand Gesture (vs. software baseline accuracy of 99\%).

\subsection{Performance, power consumption, and silicon area}\label{sec:timing}
\begin{figure}[!t] 
     \centering
     \includegraphics[width=0.99\columnwidth]{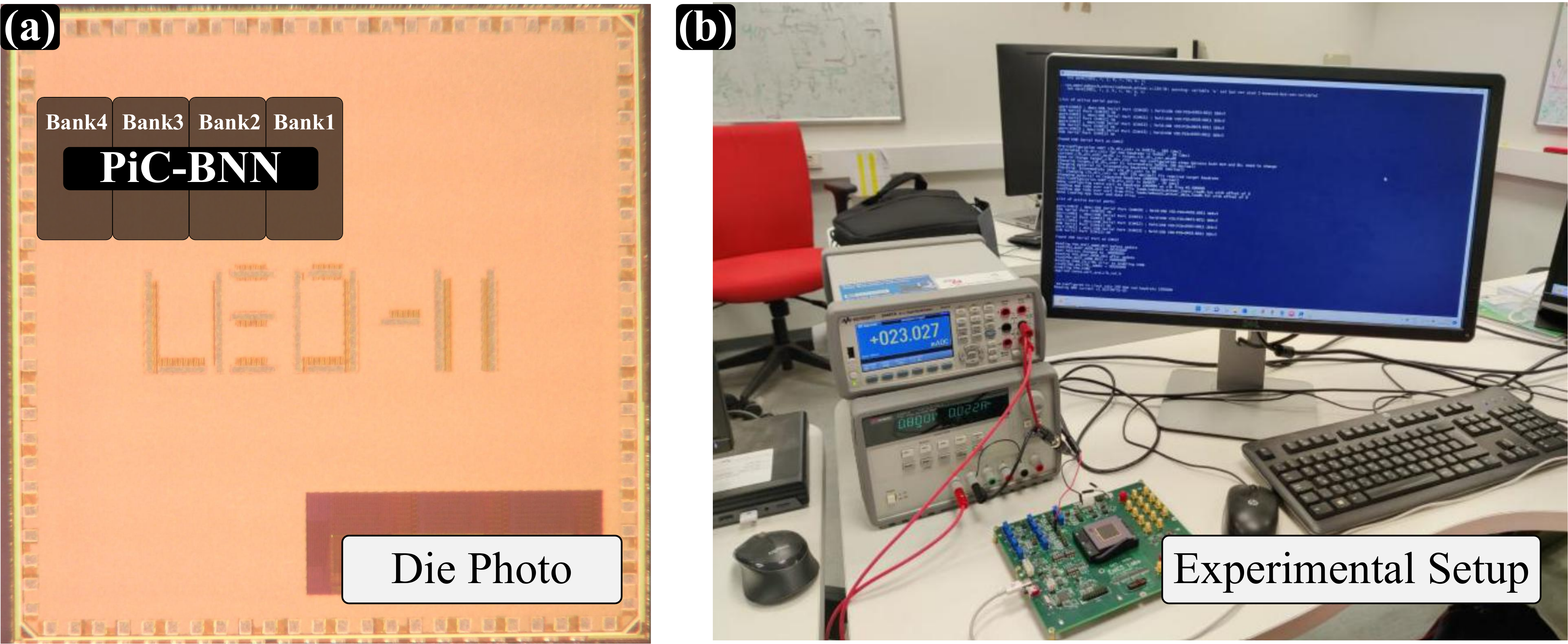} \vspace{-5mm}
     \caption{
     (a) Die photo showing PiC-BNN as a part of a research SoC ``LEO-II''. 
     (b) Evaluation setup including the PiC-BNN PCB and a workstation. 
     %Main features of the PC-CAM SoC.
     }%\vspace{-5mm}
     \label{fig:board}
\end{figure}

We use silicon measurements to evaluate the performance, area, and power consumption of PiC-BNN. 
\figref{fig:board}(a) presents the \soc micrograph with PiC-BNN marked.
\figref{fig:board}(b) depicts the PiC-BNN evaluation setup with its test (evaluation) board.

PiC-BNN was evaluated at 25\MHz, processing binary fully connected layers of up to 64$\times$2048, 128$\times$1024, or 256$\times$512 (depending on CAM banks configuration) per clock cycle. 
\black{Since tuning voltage sources is not an immediate operation, we apply batching to amortize tuning time across the processing of multiple images. 
Specifically, the same \{$V_\text{ref}, V_\text{eval}, V_\text{st}$\} combination is applied to executing the fully connected (output) layer of multiple images before re-tuning it.} 

PiC-BNN achieves inference throughput of \black{560K} inferences/s when implementing the $ 784\rightarrow128\rightarrow10$ model trained for the MNIST dataset with 33 output (fully connected) layer execution repetitions. 
The power consumption of PiC-BNN is 0.8\mW (at 25\C), therefore  \black{the effective energy efficiency of PiC-BNN is about \black{703M inferences/s/W}, or \black{up to 184 TOPs/s} without outsourcing of different BNN components to auxiliary digital processing units or software.}

\black{The overall power consumption of PiC-BNN and a RISC-V CPU that controls the \soc is about 0.3\mW.}
PiC-BNN silicon area is 0.87\mmsquared. 
\black{The overall SoC area (PiC-BNN and RISC-V CPU) is 2.38\mmsquared. }
The performance, power consumption, and silicon area figures are summarized in~\tblref{tab:soc_features}.

\begin{table}[t!]
    \centering
    \caption{Summary of PiC-BNN hardware parameters.}
    \begin{tabular}{c|c}
        \hline
         \textbf{Technology}& 65\nm CMOS\\
         \textbf{Supply Voltage}& 1.2\V\\
         \textbf{SoC Area}& 2.38\mmsquared\\
         \textbf{PiC-BNN capacity}& 128 kbit\\
         \textbf{PiC-BNN area}& 0.87\mmsquared\\
         \textbf{PiC-BNN power consumption}& 0.8\mW\\
         \textbf{PiC-BNN energy efficiency}& 184 TOPs/s\\         
         \textbf{Operating Frequency}& 25\MHz\\
        % \makecell{\textbf{Power per similarity}\\ \textbf{search}}& 1.0 - 1.3\mW\\
         \hline
    \end{tabular}
    \label{tab:soc_features}
\end{table}

\section{Conclusion}\label{sec:conclusion}
In this work, we presented PiC-BNN, an in-CAM binary neural network accelerator. 
PiC-BNN implements a true end-to-end binary neural network, replacing full precision operations, typically outsourced to host CPU or auxiliary digital units in state-of-the-art BNN solutions, by multiple executions of the output (fully connected) layer of a classification neural network. 
To ensure the results of such multiple executions differ for each run, PiC-BNN uses user-configurable voltages to adjust the Hamming distance tolerance threshold of approximate matching in its CAM arrays.  
PiC-BNN is designed and fabricated in a commercial 65\nm process, and evaluated (classification accuracy, performance, and power consumption) using silicon measurements.

%------ACKNOWLEDGEMENTS---------------%
%-------------------------------------%
\section*{Acknowledgements}
This work was supported by the Sweden-Israel Lise Meitner research collaboration
under grant number 1001569396, by the Israeli Ministry of Science, Innovation and Technology, under grants number 1001818838 and 1001702600,
by the Pazy Foundation under grant number 5100089682, and by the Italian Ministry of University and Research (MUR) under grant number
SOE 20240000022.

%----------BIBLIOGRAPHY---------------%
%-------------------------------------%
%%%%%%%%%%%%%%%%%%%%%%%%%%%%%%%%%%%%%%%%%%%%%%%%%%%%%%%%%%%%%
%    This file configures your bibliography
%        Just a bit cleaner than having a long list of bib 
%          files in your main document
%%%%%%%%%%%%%%%%%%%%%%%%%%%%%%%%%%%%%%%%%%%%%%%%%%%%%%%%%%%%%

% Start by defining the bibliography style
%    Some commonly used styles:
%       IEEEtran    - for most IEEE Transactions and Conference Proceedings
\ifacmjournal
    \bibliographystyle{ACM-Reference-Format}
\else 
    \ifmicro
        \bibliographystyle{IEEEtranS}
    \else
        \bibliographystyle{IEEEtran}
    \fi
\fi

% Now provide a list of bibliography files to include within the \bibliography command
%   These files are stored in the "bibliography" folder
%   There is an "abbreviations.bib" file which includes long and short names for common journals and confernces
%   There should be a file for papers published by each of the EnICS Staff members
%   There is a "general_bibliography" file for things we tend to cite a lot (e.g., ITRS)
%   Put your paper-specific citations in the "this_bibliography.bib" file
\bibliography{bibliography/abbreviations,
              bibliography/general_biblography,
              bibliography/teman_bibliography,
              bibliography/this_bibliography}

\end{document}